# Monolayer-to-mesoscale modulation of the optical properties in 2D CrI$_3$ mapped by hyperspectral microscopy

Marta Galbiati[1*†], Fernando Ramiro-Manzano[1,2*†], José Joaquín Pérez Grau[1], Fernando Cantos-Prieto[1], Jaume Meseguer-Sanchez[1], Ivona Kosic[1], Filippo Mione[1], Ana Pallarés Vilar[1], Andrés Cantarero[1], David Soriano[3,4], and Efrén Navarro-Moratalla[1*]

**Magnetic 2D materials hold promise to change the miniaturization paradigm of unidirectional photonic components. However, the integration of these materials in devices hinges on the accurate determination of the optical properties down to the monolayer limit, which is still missing. By using hyperspectral wide-field imaging we reveal a non-monotonic thickness dependence of the complex optical dielectric function in the archetypal magnetic 2D material CrI$_3$ extending across different length scales: onsetting at the mesoscale, peaking at the nanoscale and decreasing again down to the single layer. These results portray a modification of the electronic properties of the material and align with the layer-dependent magnetism in CrI$_3$, shedding light into the long-standing structural conundrum in this material. The unique modulation of the complex dielectric function from the monolayer up to more than 100 layers will be instrumental for understanding and manipulating the magneto-optical effects of magnetic 2D materials.**

The dependence of the physical properties of van der Waals materials with the number of layers has been the flagship of two-dimensional (2D) materials since their discovery. In general, this layer dependence gains importance upon approaching the single layer limit, where the strict confinement of electrons in a 2D lattice imposes dramatic changes in the electronic structure of the crystal, enabling the realization of new electronic states and quantum correlated phases of matter. This has opened the door for the realization of massless Dirac fermions in graphene and superconducting twisted bilayer graphene[1,2], direct-gap photoluminescence[3,4] and valley polarization in single-layer transition metal dichalcogenides[5,6], or Ising-like superconductivity in few-layer metallic transition in metal dichalcogenides[7], to name just a few prominent examples. Beyond the single layer limit, the properties of van der Waals materials change gradually until reaching the bulk properties generally within the nanoscale thickness. A much more unusual case features a continuous change of the material properties at much larger thicknesses. One of the few examples is the effect of layer number on the photoluminescence of hexagonal boron nitride[8]. However, this phenomenon arises from a variation of the bandgap and activation energies of impurities in the system and is not a consequence of the layer-dependence of the intrinsic electronic properties. Another remarkable case is chromium triiodide (CrI$_3$), one of the first layered materials to exhibit a non-zero net magnetization down to the monolayer[9], accompanied by a transition from layered antiferromagnetic in the few-layer regime to ferromagnetic in the bulk, with a crossover thickness to the bulk that is still unclear but that is unambiguously located in the mesoscale[10-12]. Seminal studies attribute this change to differences in the stacking between the layers[13] being the ferromagnetic and layered antiferromagnetic states related to the rhombohedral and monoclinic stacking, respectively. This points at a profound layer-dependent electronic effect at the mesoscale, setting an imperious need to study the evolution of the electronic properties as a function of the layer count in order to understand the underlying mechanisms giving rise to the magnetism in CrI$_3$.

An insightful and non-destructive way to study the layer-dependent evolution of the electronic properties of a layered material is to determine the complex dielectric function via light-matter interaction. Although optical ellipsometry is usually employed to extract these parameters from thin films, the small sample footprints of most mechanically exfoliated 2D materials hinder the direct application of this technique. One strategy to overcome the spatial resolution limitation is to use an optical microscope equipped with a broad-band white light source coupled to a spectrometer, where a continuous adjustment of the illumination wavelength in constrained areas of the sample allows for an accurate extraction of the dielectric function in the visible range[14]. However, the sampling speed and data statistics of point spectroscopy are generally low. Wide-field hyperspectral microscopy on the other hand has been successfully employed for the high-throughput layer-dependent characterization of bare

1   Instituto de Ciencia Molecular, Universitat de València, Calle Catedrático José Beltrán Martínez 2, 46980, Paterna, Spain.
2   Instituto de Tecnología Química, Universitat Politècnica de València - Consejo Superior de Investigaciones Científicas (UPV-CSIC), Avd. de los Naranjos s/n, 46022, Valencia, Spain.
3   Information Engineering Department, University of Pisa, Via Caruso 16, 56122, Pisa, Italy.
4   Departamento de Física Aplicada, Universidad de Alicante, 03690, San Vicente del Raspeig, Alicante, Spain.
† M.G. and F. R.-M. contributed equally to this work. * e-mail:ferraman@fis.upv.es, marta.galbiati@uv.es, efren.navarro@uv.es

2D materials and heterostructures in air[15-18]. Unfortunately, the magnetic 2D materials are very sensitive to the presence of humidity and oxygen and therefore require an appropriate isolation from ambient conditions and rapid characterization to preclude the effect of degradation, a combination of requirements that until now has not been met by the established techniques. By developing a simple encapsulation technique compatible with wide-field hyperspectral imaging, we herein circumvent the low-throughput of deterministic encapsulation and achieve high sampling frequency and data reliability in the characterization of the layer-dependent optical properties of $CrI_3$. The use of a monochrome camera with a high linear dynamic range permits a single-shot acquisition of the light intensity coming from tens of bare $CrI_3$ crystals with different layer number, providing simultaneous access to large statistics while minimizing the acquisition times. By modeling the spectral information obtained as a function of the layer number, we identify at least two crossover points of the complex optical dielectric function of $CrI_3$ in the nanoscale and the mesoscale thickness regime. The crossover in the mesoscale portrays a sizeable modification of the electronic properties of the material, which could underpin the structural preference for the monoclinic phase and provide an explanation for the layer-dependence of the magnetic properties of the material. Our theoretical results, based on first-principles calculations, provide support to the electronic origin of the evolution of the dielectric function.

$CrI_3$ flakes from the single layer to the hundreds-of-layers thickness range were obtained by mechanical exfoliation of bulk crystals on standard microscopy glass slides and encapsulated with a coverslip sealed using a thermoplastic material. Spatially resolved hyperspectral maps of different regions of the sample were then acquired by combining a sequence of optical images recorded under different monochromatic illumination spanning the full visible range (more details in the Supplementary Information SI2).

By analyzing the hyperspectral optical transmission data in selected regions of the $CrI_3$ crystals of homogeneous thickness, examined by atomic force microscopy (AFM) for their correct estimation (see SI2 for details), we obtained spectral traces of the material for different layer numbers. Figure 1 shows the transmittance spectra of flakes with a thickness ranging from 1 layer (L) to 164 L. Two absorption features are visible at about 1.95 eV and 2.7 eV, corresponding to the ligand-to-metal charge transfer absorption peaks reported for bulk $CrI_3$ and more recently for $CrI_3$ exfoliated samples[19-21]. Remarkably, both transmittance dips present a non-monotonic trend with the number of layers, i.e. a red shift when flake thickness increases from 1 L to ~ 13 L and a blue shift above ~ 50 L, thus suggesting a change in $CrI_3$ optical properties with flake thickness. It is also interesting to point out that these features remain almost unchanged for crystals thicker than ~ 100 L, hinting to a smooth saturation towards bulk values for layers thickness located at the mesoscopic scale.

After considering different approaches for the evaluation of the complex optical dielectric function ($\tilde{\varepsilon}(\omega)$

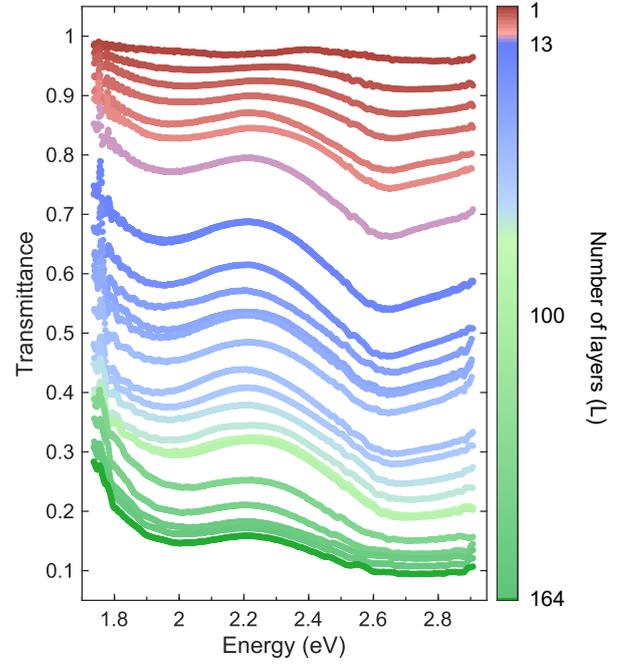

**Figure 1 | Visible range transmittance spectra of $CrI_3$ crystals with different layer number.** The data displayed was extracted from selected areas of a wide-field hyperspectral image that are found to be atomically flat by AFM. Red curves correspond to thin layers (from 1L to 13L), blue curves to layers ranging from 14L to ~100L and green curves correspond to bulk (up to 164L).

= $\varepsilon_1 - i\varepsilon_2$) from optical data (see SI3), we chose to focus on transmission measurements and large statistics analysis to achieve high data reliability. We simultaneously analysed transmittance spectra acquired over flakes with different thicknesses, including a large number of samples with ample statistics (thousands of pixels each) in every dataset, in order to limit the interdependence between $\varepsilon_1$ and $\varepsilon_2$ from a single experiment and further constrain the range of possible solutions of $CrI_3$ $\tilde{\varepsilon}(\omega)$.

To extract the complex dielectric function from the transmittance data, we consider our experimental system as a basic Fabry-Pérot cavity formed by a stack of 3 layers and we assume that the dielectric function of $CrI_3$ follows a modified Lorentzian oscillator model[22]:

$$\tilde{\varepsilon}(\omega) = \tilde{\varepsilon}_\infty + \sum_{n=1}^{2} \frac{\omega_p^2 + 2i\omega\Gamma_n}{\omega_n^2 - \omega^2 + 2i\omega\gamma_n}$$

This model is composed by two oscillators ($n$ = 1,2), where $\tilde{\varepsilon}_\infty$ is the permitivity for infinite optical frequencies, $\omega_n$, $\omega_p$, and $\omega$ represent the resonant, plasma and photon frequency, respectively, and $\gamma_n$ and $\Gamma_n$ are damping related parameters. Figure 2 shows the real and imaginary parts of $\tilde{\varepsilon}(\omega)$ and their evolution with flake thickness calculated by simultaneous fitting of all the transmittance spectra shown in Figure 1 (see SI3 for details). The dielectric function of $CrI_3$ monolayer is found to be significantly different from the rest of the layers. This is in line with the transmittance spectra experimentally measured in the monolayer, where absorption features are significantly shifted compared to the bilayer. Besides the discontinuity found for the monolayer, we find two additional critical points at about 13 L (~8.9 nm) and 100 L (~68.7 nm), where $\varepsilon_2$ intensity increases and peaks red shift (< 13 L), then decreases and peaks blue shift (> 13L), until the dielectric function starts to asymptotically saturate toward the bulk value above ~100 L. This delimits three different thickness domains: (i) few layer, (ii) multilayer and (iii) bulk. These changes are also clearly illustrated in Figure 3 which shows the evolution with layer thickness of $\varepsilon_1$ and $\varepsilon_2$ at 2.66 eV extracted from the fitting process (Figure 3a), and the evolution of the experimental absorption feature minima of the transmittance spectra (Figure 3b). Hyperspectral analysis also gives the possibility to spatially resolve the optical properties of the 2D material. As such, Figure 3c displays the spatially resolved wide field image of the higher-energy transmittance dip of $CrI_3$ flakes for different thicknesses (see SI4 for calculation details) where the dip in energy of the crossover from a few layers to bulk is also clearly visible. These experimental observations hence point at profound changes in the optical properties of $CrI_3$ in these three different thickness domains.

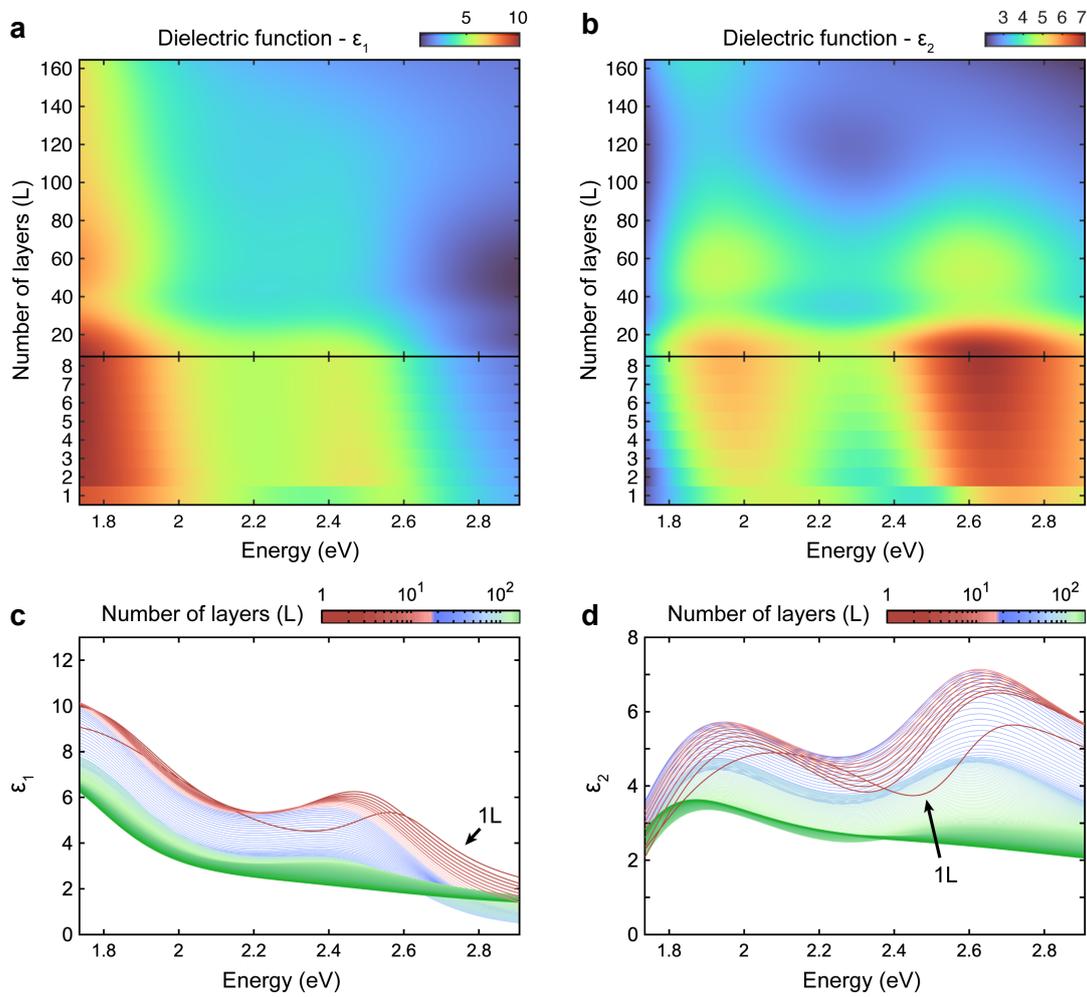

**Figure 2 | Evolution of the dielectric functions in CrI$_3$ as a function of the layer number.** The calculated real ($\varepsilon_1$) and imaginary ($\varepsilon_2$) parts of are shown in panels (a) and (b), respectively. (c) and (d) have been also plotted as a function of the excitation energy for crystals of different thicknesses (these are line cuts of the image plots shown in (a) and (b)).

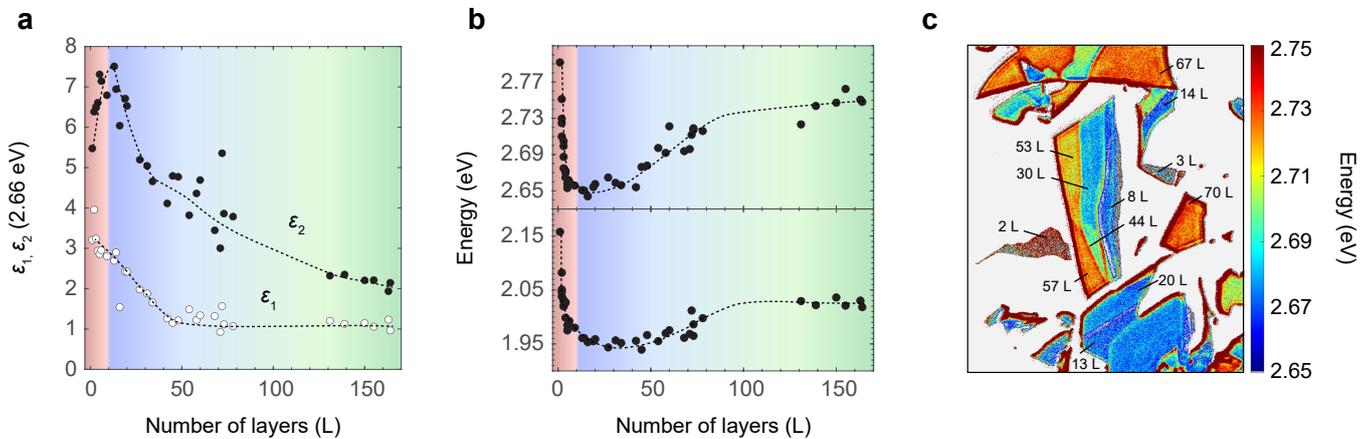

**Figure 3 | Identification of different thickness regimes in CrI$_3$.** The few-layer, multilayer and bulk thickness regimes are depicted according to the layer dependence of $\varepsilon_1$ and $\varepsilon_2$ at 2.66 eV (a) and the low (bottom) and high (top) energy transmittance minima (b). Markers display experimental data while the dotted lines are guides to the eye. (c) Spatially resolved wide field image of the high energy transmittance dip position calculated from hyperspectral images of CrI$_3$ flakes with different thicknesses. The energy crossover of dip position when increasing the number of layers is clearly visible, highlighting the different thickness regimes.

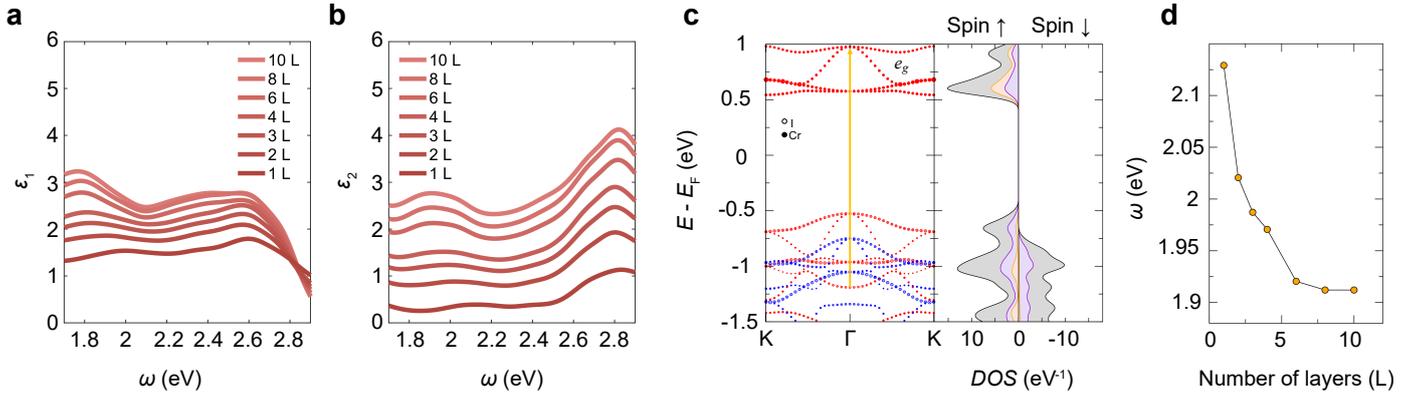

**Figure 4 | Theory calculations of the layer-dependent electronic and optical properties of $CrI_3$.** (a-b) Evolution of $\varepsilon_1$ and $\varepsilon_2$ of $CrI_3$ with increasing number of layers calculated in the z-direction perpendicular to the layers. (c) Projected band structure and DOS of monolayer $CrI_3$. Filled and empty circles represent I p-orbitals and Cr d-orbitals respectively. The size of the circles is the weight of the orbital wavefunction in each band. Red and blue stand for spin up and down respectively. The yellow arrow shows the most probable transition responsible for the low-energy peak in the dielectric function. In (d), we show the evolution of the low-energy peak in the imaginary complex dielectric function with increasing number of layers.

The evolution of the dielectric function with the layer number has been previously reported in other 2D materials such as transition metal dichalcogenides[23], $In_2Se_3$[24], $PdSe_2$[25] and a non-monotonic behaviour, consistent with our work, was observed in $MoS_2$ at the nanoscale[26]. However, in all these cases this behaviour has been reported only in the nanoscale while this is the first time that a modulation of the dielectric function has been experimentally observed at the mesoscale range of thickness, allowing for a continuous modulation of the optical properties from 1 to more than 100 layers.

To gain further insight into the possible origins of this mesoscopic transition, we carried out first-principles calculations on few-layer $CrI_3$, up to 10 L (see SI5 for details). In the real and imaginary parts of the computed dielectric function (Figure 4a-b), we identify 2 peaks located at 2.1 eV and 2.8 eV, in good agreement with the experimental results. A look at the band structure and DOS of monolayer $CrI_3$ (left panel in Figure 4c) confirms that these peaks can be ascribed to metal-to-ligand charge transfer processes between the p-orbitals of the ligands, localized in the last occupied bands, and the empty $d_{x^2-y^2}$ and $d_{z^2}$ orbitals of the chromium atoms localized in the $e_g$ set of conduction bands. In the right panel of Figure 4c, we plot the evolution of the peak at 2.1 eV with increasing number of layers. We observe a strong red shift from 2.1 eV to 1.9 eV in agreement with experimental data and indicating a clear connection between the number of layers and the electronic structure in few-layer $CrI_3$. The calculations also show a tendency towards saturation of the value of the complex dielectric function when increasing the thickness from 1 L to 10 L, which confirms the electronic origin of the layer-dependence. The discrepancies in the absolute value of the complex dielectric functions extracted from the calculations and the experimental ones may originate from subtle differences of the inter-layer distance at different temperatures and layer numbers, which have a strong effect on the polarizability of the wave-functions in the direction perpendicular to the layers (see SI5 for more details). This effect also suggests that small changes in the crystal structure could eventually modify the intensity of the complex dielectric function. Indeed, our calculations indicate that at 0 K the interlayer distance in bulk layered antiferromagnetic $CrI_3$ increases up to 0.2 Å compared to few layer samples (see Table S2) which may lead to a reduction of the polarizability and the dielectric function. This result hence highlights the dipolar and long-range nature of the van der Waals interactions which can be one of the possible explanations for the observed mesoscopic crossover of the complex dielectric function experimentally observed in the different thickness regimes.

The saturation of the value of the complex dielectric function at the mesoscale is well aligned with the critical crystal thickness at which the low-temperature magnetic properties of $CrI_3$ change from antiferromagnetic to ferromagnetic. Although this critical thickness has not yet been unambiguously determined, most works report on the layered antiferromagnetic state persisting in crystals up to 50 nm[27,28]. Both the magnetic and the optical crossover thickness to the bulk regime being in the same range of thickness is a strong indication of a connection between both phenomena. Our findings, supported by the theoretical calculations, point at an electronic origin of these effects, with an evolution of the electronic structure with the layer number onsetting below 100 L (68.7 nm) as revealed by the non-monotonic modulation of the dielectric function as the crystal is thinned down. These results will contribute to shed light into the open structural conundrum of the layer-dependent phase diagram of $CrI_3$.

Our results demonstrate that hyperspectral transmission microscopy is instrumental to study the layer-dependent evolution of the electronic properties of air-sensitive layered materials through the determination of its complex dielectric function. The possibility to obtain in a single acquisition the light intensity coming from tens of bare $CrI_3$ crystals with different layer number guarantees high throughput statistics and allows for a robust and simultaneous determination of both the real and imaginary layer-dependent components of the dielectric function. This approach reveals at least two crossover points of the electronic properties of $CrI_3$, depicting changes of the trends of the complex dielectric function both in the nano- and the mesoscale thickness regimes. The significantly wide span of the continuous layer number modulation of the optical and electronic structure, covering more than 100 layers, will be pivotal to enable the fine tuning of the optical properties of magnetic 2D materials for the development of new magnetic 2D-based devices, such as unidirectional miniaturized photonic components.

*Acknowledgments*

The project that gave rise to these results received the financial support of a fellowship from "la Caixa" Foundation (ID 100010434, fellowship codes LCF/BQ/PR21/11840011 and LCF/BQ/DI22/11940022) and the grant PID2020-118938GA-100 from the Spanish Ministerio de Ciencia e Innovación (MICINN). ENM acknowledges the European Research Council (ERC) under the Horizon 2020 research and innovation program (ERC StG, grant agreement No. 803092) and to the MICINN for financial support from the Ramon y Cajal program (Grant No. RYC2018-024736-I). FCP also acknowledges the MICINN for the FPU program (Grant No. FPU17/01587). This work was also supported by the Spanish Unidad de Excelencia "María de Maeztu" (CEX2019-000919-M) and is part of the Advanced Materials programme supported by MICIN with funding from European Union NextGenerationEU (PRTR-C17.I1) and by Generalitat Valenciana.

*Methods*

**Crystal growth and isolation of few-layer crystals:** High quality $CrI_3$ crystals were grown by chemical vapor transport and thoroughly characterized by X-ray diffraction (XRD), Energy-dispersive X-ray spectroscopy (EDAX) and Raman spectroscopy to verify their pristine quality (see Supporting Information SI1 for details). Atomically thin layers were obtained by mechanical exfoliation of bulk crystals using a scotch tape technique and deposited over a transparent glass slide to perform transmittance measurements. Selected flakes were first selected by optical microscopy screening and successively characterized by atomic force microscopy to accurately determine their thickness. The whole process and characterization were carried out inside an Ar-filled glove box where the $O_2$ and $H_2O$ levels were maintained below 0.5 ppm to avoid material degradation.

**Optical transmission experiment:** Once the thicknesses of all flakes were determined, the sample was encapsulated by gluing on the glass slide substrate a coverslip with a thermoplastic material and exposed to the air avoiding degradation. Hyperspectral transmittance images were acquired under monochromatic light ranging from 430 nm to 720 nm in an upright transmittance microscope equipped with a high-resolution CMOS monochromatic camera (Hamamatsu Orca Fusion) and a homemade grating monochromator for the incidence light, for accurate quantitative transmittance measurements.

**Extraction of the optical dielectric function:** The transmittance spectra have been fitted simultaneously by employing a multilayer model whose complex coefficients of reflectance and transmittance are described by Fresnel equations. Their complex parameters (refractive index and extinction coefficient) have been calculated from the optical dielectric function (Equation 1) assuming a relative permeability of $\mu_r$=1 (at room temperature). To fit the model simultaneously with all the experimental data (Figure 1), we have considered a spline for modelling the layer dependence of the parameters of Equation 1. In particular, a faithful data fit could be reached employing two or three cubic polynomials (C1 continuity) excluding or including the monolayer, respectively.

**Density functional theory:** The electronic structure calculations and geometry relaxations have been carried out using Quantum-Espresso ab initio package. Each structure has been relaxed using the Grimme-D3 van der Waals correction until all the forces acting on atoms were lower than $10^{-3}$ Ry/bohr. For the relaxation, we have used projector augmented waves (PAW) pseudopotentials within the generalized gradient approximation (GGA) and the Perdew-Burke-Ernzerhof (PBE) exchange-correlation potential, with a 4x4x1 k-point grid. The dielectric functions were obtained using the epsilon.x code, which is part of the Quantum-Espresso package. The calculations were carried out at the independent particle (IP) level and non-local contributions from pseudopotentials are neglected. For the optical properties, the electronic structure calculations are carried out using the DFT+U+J scheme with U = 4.1 and J = 0.6 eV norm-conserving GGA-PBE pseudopotentials from the PseudoDojo library, with energy cutoffs 80 and 320 Ry for the wavefunctions and charge density, respectively. In our case, the dielectric functions converged for a 8x8x1 k-point grid.

# Supporting information

# Monolayer-to-mesoscale modulation of the optical properties in 2D CrI$_3$ mapped by hyperspectral microscopy

Marta Galbiati[1*†], Fernando Ramiro-Manzano[1,2*†], José Joaquín Pérez Grau[1], Fernando Cantos-Prieto[1], Jaume Meseguer-Sanchez[1], Ivona Kosic[1], Filippo Mione[1], Ana Pallarés Vilar[1], Andrés Cantarero[1], David Soriano[3,4], and Efrén Navarro-Moratalla[1*]

## Contents

1  Instituto de Ciencia Molecular, Universitat de València, Calle Catedrático José Beltrán Martínez 2, 46980, Paterna, Spain.
2  Instituto de Tecnología Química, Universitat Politècnica de València - Consejo Superior de Investigaciones Científicas (UPV-CSIC), Avd. de los Naranjos s/n, 46022, Valencia, Spain.
3  Information Engineering Department, University of Pisa, Via Caruso 16, 56122, Pisa, Italy.
4  Departamento de Física Aplicada, Universidad de Alicante, 03690, San Vicente del Raspeig, Alicante, Spain.
† M.G. and F. R.-M. contributed equally to this work. * e-mail:ferraman@fis.upv.es, marta.galbiati@uv.es, efren.navarro@uv.es

# 1. Materials characterization and sample preparation

## 1.1. Bulk CrI$_3$ crystal characterization

*Powder X-ray diffraction (PXRD):*

The crystal structure of CrI$_3$ was characterized by powder X-ray diffraction (PXRD). Figure S1 shows the XRD of the of CrI$_3$ single crystals taken employing PANalytical Empyrean X-ray diffractometer. X-Rays diffraction analysis was performed on sample of single crystalline CrI$_3$ by loading the material into a capillary and sealing it inside the glove box. The powder pattern of sample is consistent with the monoclinic AlCl$_3$-type structure (C2/m) reported for CrI$_3$. In the XRD spectrum, no evidence of any other phases was detected indicating that the product is of high purity.

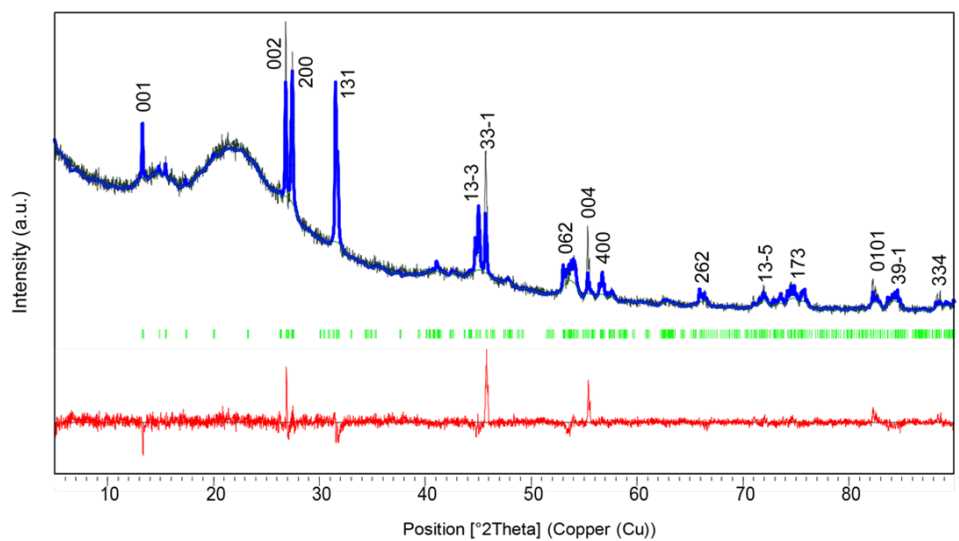

**Figure S1 | XRD pattern and unit cell refinement of a representative CrI$_3$ single crystal.** The continuous black line shows the experimental pattern, the blue line shows the Rietveld refined total intensity and the red trace depicts the difference. The green marks indicate the refined reflections. A selection of the most prominent reflections are labeled using their corresponding Miller indices.

*Energy Dispersive Analysis of X-ray (EDAX):*

The chemical composition of as-grown CrI$_3$ single crystals was determined by Energy Dispersive Analysis of X-ray (EDAX) technique using Hitachi S4800 for confirming stoichiometry. The obtained spectrum is shown in Figure S2. The atomic and weight % obtained using EDAX is tabulated in Table S1. The data clearly states that the as-grown CVT single crystals have chemical composition of CrI$_3$ and they are free from any impurity.

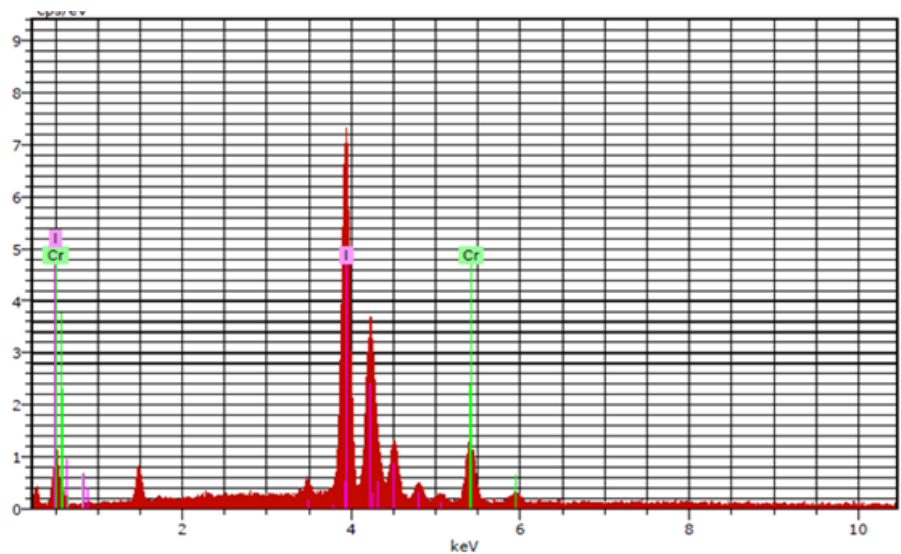

**Figure S2 | EDAX spectrum of the as-grown CrI$_3$ single crystals.**

| Element | Weight % | Atomic % |
|---------|----------|----------|
| Cr | 11.73 | 24.49 |
| I | 88.27 | 75.51 |

**Table S1 | EDAX spectrum of the as-grown CrI$_3$ single crystals.**

*Raman spectroscopy:*

CrI$_3$ bulk flakes have been characterized by Raman spectroscopy after their mechanical exfoliation to verify their pristine quality after encapsulation with the glass cover (see section 1.2 below). Raman characterization has been performed using a Horiba LabRAM HR Evolution Raman spectrometer under a 532 nm excitation laser. Results in Figure S3 show peaks at frequencies of 78, 100–110, 128, and 234 cm$^{-1}$ in agreement to that previously reported.

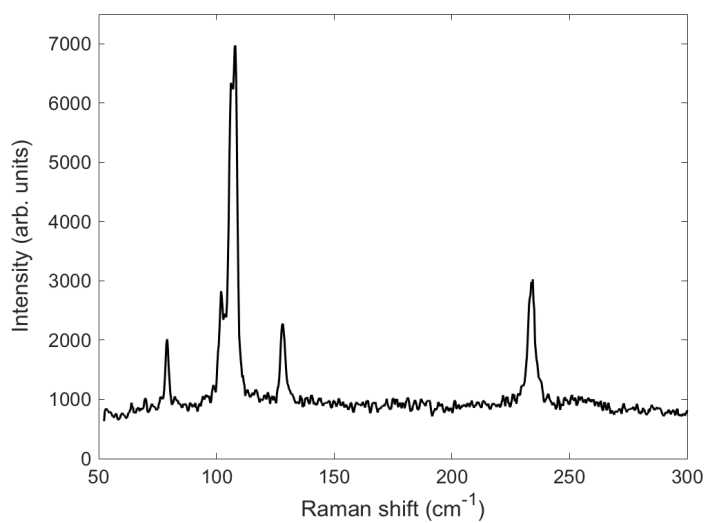

**Figure S3 | Raman spectrum of bulk CrI$_3$ flake.** The flake under observation was well within the bulk thickness regime described in the main text.

## 1.2. Preparation of samples for transmission experiment

For our transparent substrates we use standard microscope glass slides thoroughly rinsed to obtain very clean and low roughness surfaces. Glass slides were soaked in a freshly prepared solution of NH$_4$OH : H$_2$O$_2$ (1 : 1) and sonicated for 8 min. Next, they were rinsed with Milli-Q water, sonicated 5 min in Milli-Q water and dried under a nitrogen stream. Clean slides were introduced inside the glove box and heated above 100$^{\circ}$C for at least 30 min before deposition of the 2D material, in order to remove any possible H$_2$O traces from the substrate surface.

Atomically thin CrI$_3$ layers were obtained by mechanical exfoliation of bulk CrI$_3$ crystal, grown by chemical vapor transport, using a scotch tape technique and deposited over the glass slide. An optical microscope was then used to screen the sample and select regions of interest with abundant flakes. Selected layers were characterized by atomic force microscopy to determine their thickness. Remarkably, the surface of the glass slides was observed to be perfectly clean with a low roughness (RMS ~ 0.23 nm) which allows accurate flakes thickness measurement down to the monolayer.

The whole exfoliation and characterization processes were carried out inside an Ar-filled glove box where O$_2$ and H$_2$O levels were maintained below 0.5 ppm to avoid material degradation. Once flakes characterized, sample was sealed with a thin glass cover slide (100 µm thick) to prevent CrI$_3$ layers degradation when exposed to the air. The samples were finally extracted from the glove box to proceed with the wide field hyperspectral experiment.

## 2. Wide-field hyperspectral setup for air-sensitive materials

A wide field hyperspectral setup has been implemented by modifying a standard optical transmittance microscope as schematically shown in Figure S4. White light coming from a high intensity LED phosphor light source enters a homemade monochromator, which allows selecting the excitation wavelength with a bandwidth of ~1 nm. Monochromatic light is then connected to the optical microscope through an optical fiber and is used to irradiate the transparent samples in transmittance mode. For the hyperspectral analysis, transmission images of a region of interest of the sample are acquired sweeping the wavelength from 430 nm (2.9 eV) to 720 nm (1.7 eV) with a step of 1 nm. To achieve this, the optical microscope is equipped with a monochrome CMOS camera (Hamamatsu Orca Fusion), conveniently chosen for its low noise and high linear range, which allows precise quantitative analysis of the images even for thin layers and in the full visible range. Data acquisition is performed in complete dark conditions, in order to avoid any source of external light, which could result in reduced data reproducibility. With an exposure time of about 300 ms per image, a complete hyperspectral image takes less than 10 minutes.

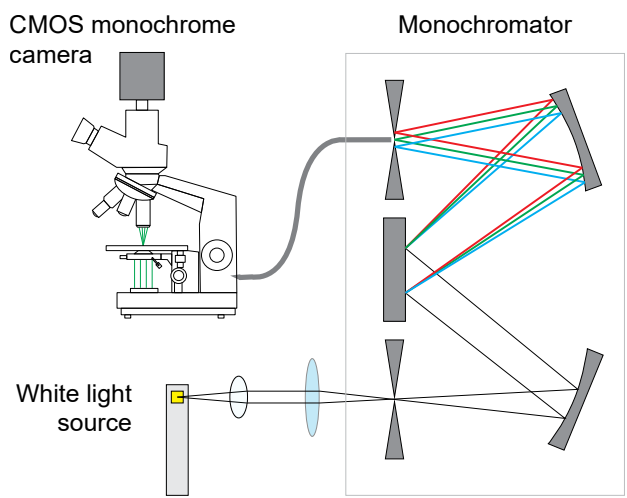

**Figure S4 | Schematic of the wide-field hyperspectral microscope setup.** Solid straight lines depict selected light ray traces. The drawing is not to scale.

Figure S5 shows transmission images of $CrI_3$ atomically thin crystals down to the monolayer under different wavelengths, with a contrast that ranges from almost transparent to very dark for increasing thickness. Figure S5 also demonstrates the high throughput of this imaging technique, which permits exploring the layer-dependent optical properties in the visible range from the single layer to the hundreds-of-layers range with ample statistics coming from just a single frame of hyperspectral images. To accurately determine the number of layers of each flake we employed atomic force microscopy (AFM). As shown in Figure S5e,f, the topographic image of a $CrI_3$ single layer on a glass slide depicts a thickness of ~1.1 nm which is in agreement, within the experimental uncertainty, with the interlayer space extracted from the lattice parameter of the bulk structure [1], and demonstrates the atomically-flat quality of the micrometer-size samples obtained down to the single layer limit. It is also worth noticing that the root-mean-square (RMS) roughness measured over the commercial glass slide is only ~ 0.23 nm, comparable to that obtained on standard $Si/SiO_2$ wafers, hence making commercial glass slides a convenient transparent substrate even for the investigation of the thinnest layers.

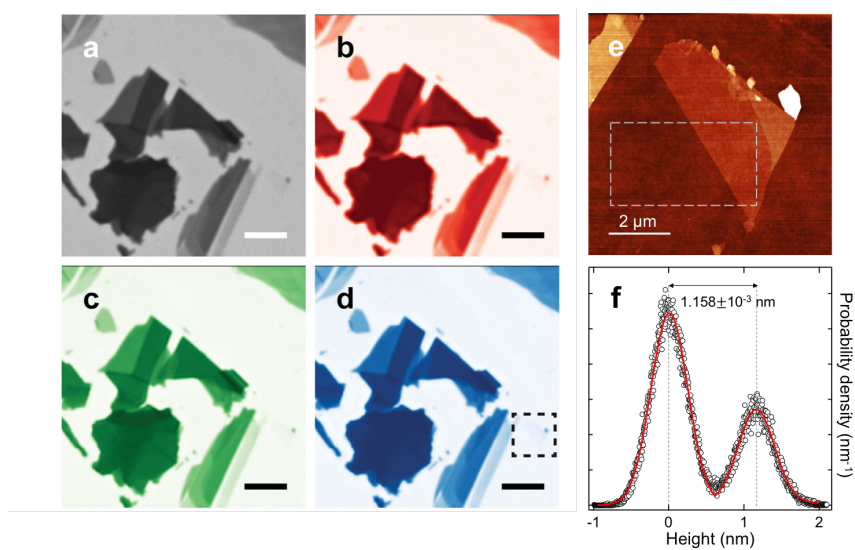

**Figure S5 | Few-layer $CrI_3$ crystals deposited on standard glass slides.** (a-d) Optical transmission micrographs of a representative sample, showing different layer number plateaus illuminated with white, 630 nm, 532 nm and 460 nm light respectively. Scale bars in the optical micrographs are 10 µm long. (e) Atomic force microscopy topographic image of single layer $CrI_3$ located inside the dashed box drawn in panel (d), with the corresponding probability density distribution shown in panel (f), which allows to estimate a flake thickness of 1.15 ± 0.001 nm, compatible with a monolayer within experimental uncertainty. Probability density has been calculated in the region marked with a dotted line in panel (e).

The collected hyperspectral data are arranged in a 3D matrix where each image pixel (x and y spatial coordinates) is associated to all the swept wavelengths. To reduce signal to noise ratio in data analysis we select a region with homogeneous thickness on the analyzed flake (blue mask) and a second neighboring region on the substrate (green mask). Typical flake/substrate regions are displayed in Figure S6a. A dark background frame is then subtracted and the pixels intensities are averaged for each region ($I_{flake}$ and $I_{substrate}$). Quantitative transmittance is calculated as $T(E) = I_{flake}(E)/I_{substrate}(E)$. Figure S6c shows examples of typical transmittance spectra for different $CrI_3$ flake thicknesses. Good reproducibility of the transmittance spectra for different flakes with equal thickness was verified in order to ensure a reliable layer-dependent analysis. Reflection data showed to be strongly affected by the choice of regions and by spurious reflections, not complying with our data reproducibility requirement and resulting in a high dispersion of spectral traces for different flakes with equal thickness. Reflection data was consequently excluded from our analysis.

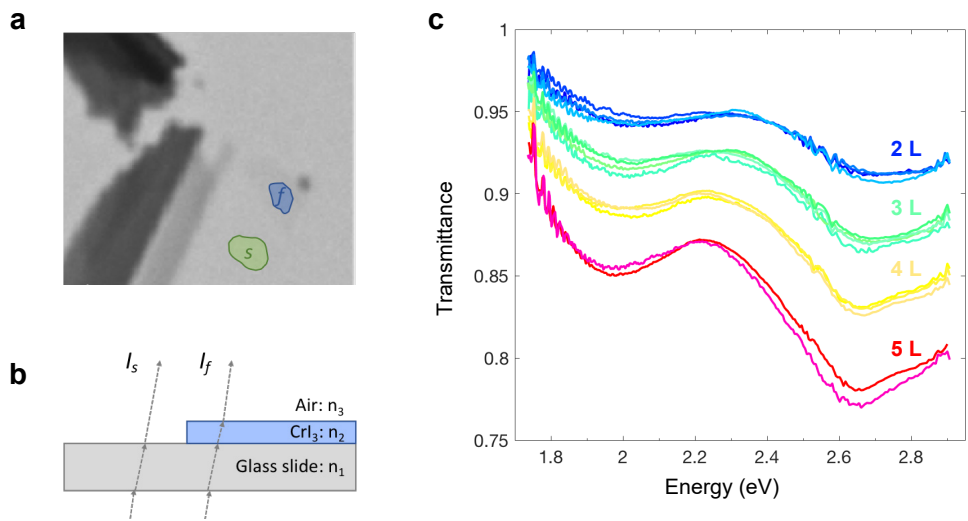

**Figure S6 | Transmission data processing and representative results.** (a) Optical transmission image of $CrI_3$ flakes deposited over a glass slide. In order to analyze transmittance spectra a zone of the flake with the same thickness is selected over the image (area labeled with an f), and a zone over the substrate just nearby the flake (area labeled with an s). Transmission intensities are then averaged over the selected zones (If and Is) for each image acquired in a wavelength range of 430 - 720 nm. Finally, transmittance is calculated as $T = I_f/I_s$ for each swept wavelength (energy). (b) Schematic of the simplified system where monochromatic light is coming from the back of the sample in transmittance mode and is transmitted through the glass slide and the $CrI_3$ layers before being collected by the camera. (c) Examples of transmittance spectra collected for different $CrI_3$ layers thickness. To ensure data reliability good reproducibility of transmittance spectra over different flakes with the same thickness has been verified as it can be observed in this graph.

## 3. Optical model

In previous works, the reflectance of 2D materials deposited on solid substrates has been modeled using the Fresnel equations [1–4]. In this regard, the real ($\varepsilon_1$) and imaginary ($\varepsilon_2$) part of the optical dielectric function of individual flakes of the materials are derived from a single reflectance experiment using a Kramers-Kronig (KK)-constrained analysis [5]. This approach has been used, for instance, to find the dielectric function of TMDC monolayers deposited on fused silica substrates modeling $\tilde{\varepsilon}(\omega)$ as a superposition of Lorentzian oscillators, whose complex equations fulfill the KK relations.[3] Similarly, KK relations have been used to calculate the phase of the amplitude reflection coefficient $\theta$ to extract the refractive index and extinction coefficient of bulk $CrI_3$ in the visible range [1], or in the terahertz to near-infrared region [6]. However, this approach has the important limitation that, for an accurate determination of $\tilde{\varepsilon}(\omega)$, KK relations require a wide energy range (ideally from 0 to ∞ while optical contrast experiments are usually limited to the small range of visible light. To address this issue, a possible solution has been to make use of literature data to extrapolate high and low energy limits. While we initially applied this same approach, we realized that for our spectral range, the extracted phase ($\theta$) showed a remarkable dependence with the reference data. As a result, it was difficult to reach consistent solutions over different spectra.

An alternative technique to determine the optical dielectric function is to acquire both reflectance and transmittance spectra for solving a system of two independent equations. In this way, both $\varepsilon_1$ and $\varepsilon_2$ can be unequivocally obtained. At this aim, we attempted to simultaneously analyze the reflectance spectra of $CrI_3$ over two different substrates: $SiO_2$/Si and a glass slide, successively trying to strengthen our analysis by additionally relating them to transmittance measurements performed on the same $CrI_3$/glass slide sample. However, this extended analysis led us to the following conclusions: firstly, reflection spectra introduce a high experimental error, which hampered finding a solution for the computed $\tilde{\varepsilon}(\omega)$ of the 2D material. While $SiO_2$/Si substrate interference strongly masked the small absorption peaks of $CrI_3$ layers, external reflections due to optical elements, very difficult to get rid of, were observed on the transparent glass substrate, resulting in a poor reproducibility of reflectance data and subsequent inconsistent analysis. Secondly, we verified that $\varepsilon_2$ dominates both the transmittance (see Figure S8) and reflectance spectra [3], leading to ambiguity of the solution of the equation due to the high experimental uncertainty.

We thus focused on transmission measurements as a source of high data reliability. To limit the interdependence between $\varepsilon_1$ and $\varepsilon_2$ from a single experiment we simultaneously analyzed transmittance spectra acquired over flakes with different thicknesses, including a large number of samples with ample statistics (thousands of pixels each) in every dataset. This approach allows to further constrain the range of possible solutions of $CrI_3$ $\tilde{\varepsilon}(\omega)$, hence increasing the consistency of our analysis.

For the transmittance analysis we describe our system as a simple Fabry-Perot cavity composed of three layers as shown in Figure S6b: (1) glass slide / (2) $CrI_3$ / (3) air. Other contributions such as the presence of a glass cover or the rest of optical elements cancel each other in the transmittance model. The energy dependent complex refractive index for each layer is written as:

$$\tilde{n}_1(\omega) = n_{glass}(\omega) - ik_{glass}(\omega); \; \tilde{n}_2(\omega) = n_{2D}(\omega) - ik_{2D}(\omega); \; n_3(\omega) = 1.$$

To calculate the complex refractive index of $CrI_3$ ($n_2(\omega)$), we assume $CrI_3$ dielectric function to follow a modified Lorentzian oscillatory model:

$$\tilde{\varepsilon}(\omega) = \tilde{\varepsilon}_\infty + \sum_{n=1}^{2} \frac{\omega_p^2 + 2i\omega\Gamma_n}{\omega_n^2 - \omega^2 + 2i\omega\gamma_n}$$

composed by two oscillators ($n$ = 1, 2), where $\omega_n$, $\omega_p$, $\omega$ represent the resonant, plasma and scan frequencies respectively, $\tilde{\varepsilon}_\infty$ is assumed to be 1, and $\gamma_n$ and $\Gamma_n$ correspond to damping related parameters. We then extract $n_2$ from $n_2(\omega) \approx \sqrt{\tilde{\varepsilon}(\omega)}$.

Figure S7 shows examples of transmittance spectra (black lines) and fit (red lines) performed with our model on $CrI_3$ samples with different thickness. A very good agreement between experimental data and theoretical fit can be observed.

Figure S8 shows maps of solutions of the model around the energy values of the transmittance dips (1.95 and 2.66 eV). These maps were generated by plotting the difference, in absolute value, between the experimental transmittance signal and different theoretical solutions. These model solutions are simulated by scanning different combinations of $\varepsilon_1$ and $\varepsilon_2$. The zeros of this subtraction (black/dark region) correspond to possible solutions. Given that these values form quasi horizontal lines, the solution is dominated by a small and limited range of $\varepsilon_2$ values in contrast to a larger dispersion of $\varepsilon_1$. Indeed, a certain deviation of $\varepsilon_1$ results in a slight perturbation of the highly reliable $\varepsilon_2$ value. In any case, the solutions of the maps show a small curvature due to the multilayer nature of the system. Consequently, this shape changes with the sample thickness. Importantly, in our experiments the layer complexity was streamlined to the bare minimum, i.e. consisting of a single layer substrate holding the sample, and as a result, the Figure S8 shows simple and almost flat solution curves. The white marker in each map indicates the solution captured by the fitting procedure. Note that the fitting procedure not only considers a single spectral energy (as shown in Figure S8), but it considers the full spectroscopic experimental data range (Figure S7) for obtaining the solution Note that our final fitting procedure not only considers a single spectral energy (as shown in Figure S8), but it considers the full spectroscopic experimental data range (Figure S7) for obtaining a complete solution as a function of the layer number.

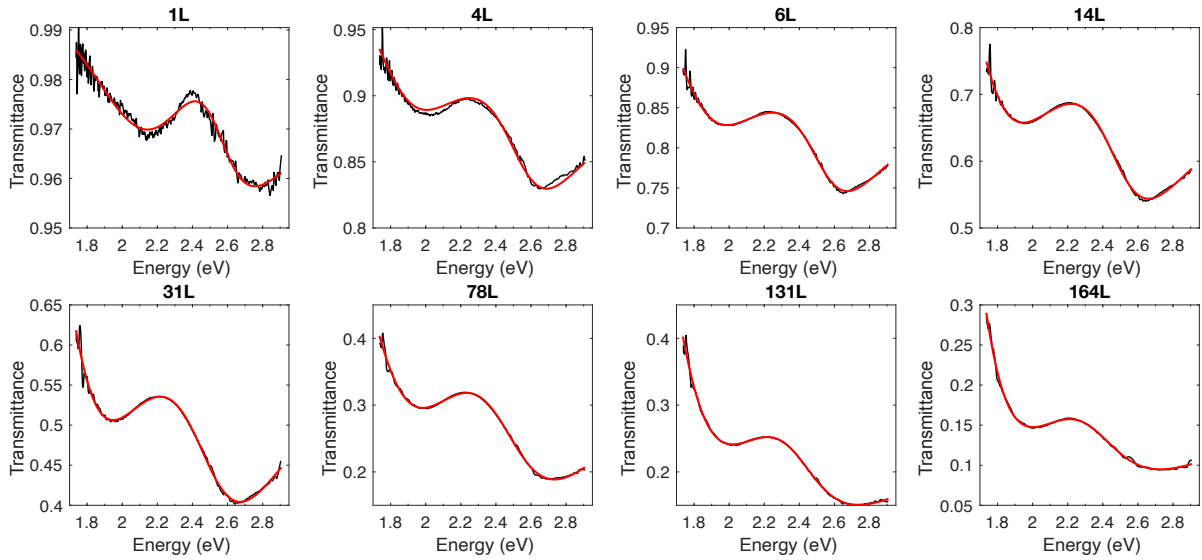

**Figure S7 | Examples of transmittance spectra experimentally measured over CrI$_3$ layers of different thickness (black curves) and their model fit (red curves).** The results plotted were obtained by using the oscillatory model for different number of layers.

To further increase consistency of our data analysis, we perform a simultaneous fit using the previously described model with > 40 transmittance spectra of CrI$_3$ layers with thickness ranging from 1 L to 164 L. Although each spectrum represents a different scenario, they are interconnected by shared parameters with a defined tendency. We rely on the fits performed individually on each spectrum for the different layer thicknesses in order to gain a first idea on the fitting parameters evolution with the number of layers (Figure S9). Then, we fit parameters tendency with respect to layer thicknesses with a spline consisting in 3 cubic polynomials defining two C1 intersection points (or knots) at about 1L and 46 L. Our model allows to fits accurately to all transmittance spectra while interconnecting fitting parameters allows to highly constrain the number of possible solutions, hence increasing the accuracy of our analysis (Figure 2). It is important to point out that it was not possible to attain a good fit of all transmittance spectra without defining the spline intersection points (at about 1L and 46L). This, again, supports the idea of three thickness regimes with different optical properties: the so-called thin, multilayer and bulk regions as discussed in the main text.

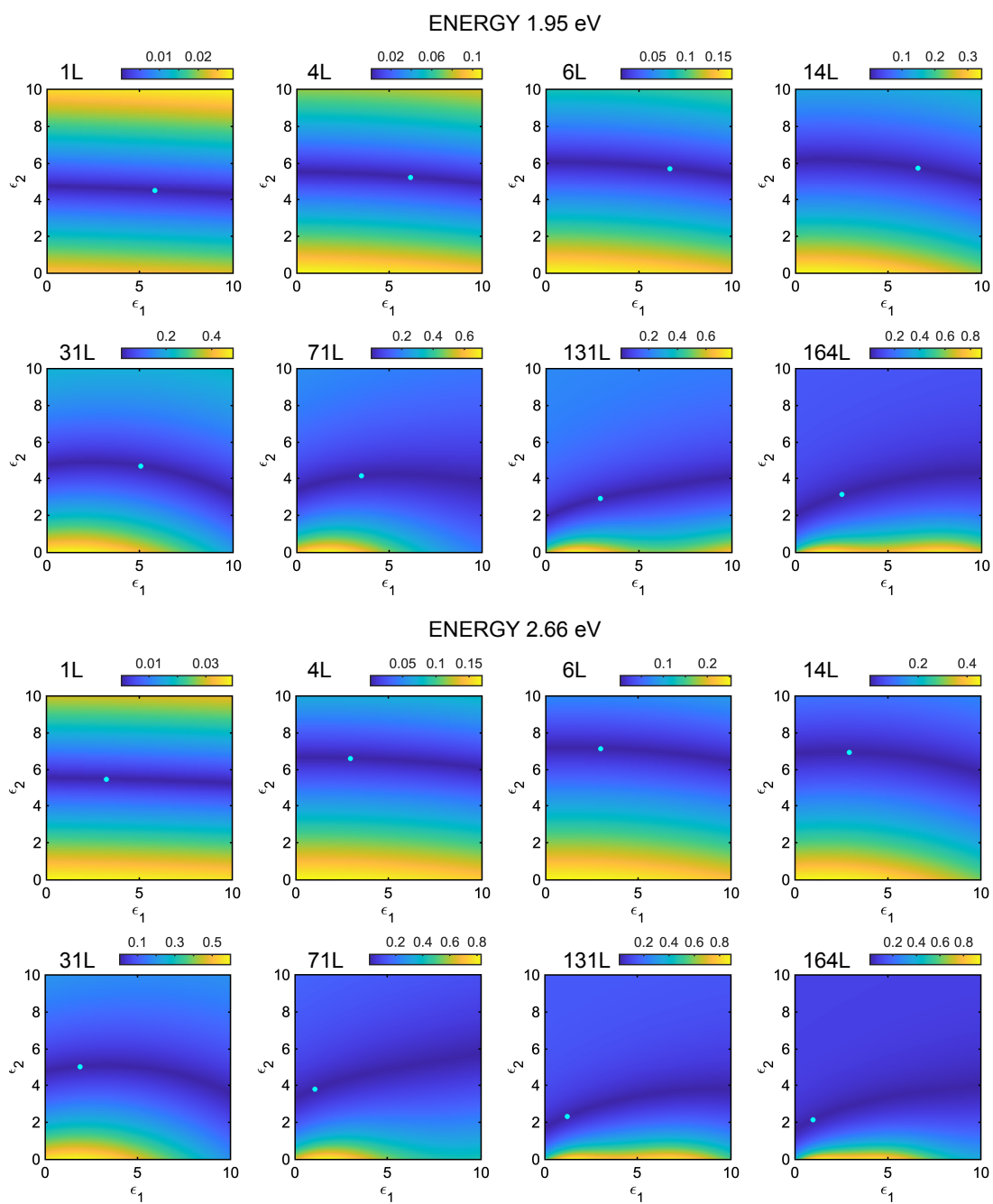

**Figure S8 | Map of solutions of the model computed at 1.95 eV and 2.66 eV.** Black regions correspond to the possible solutions (zeros) as a function of $\varepsilon_1$ and $\varepsilon_2$. White dots represent individual solutions extracted from the fitting process.

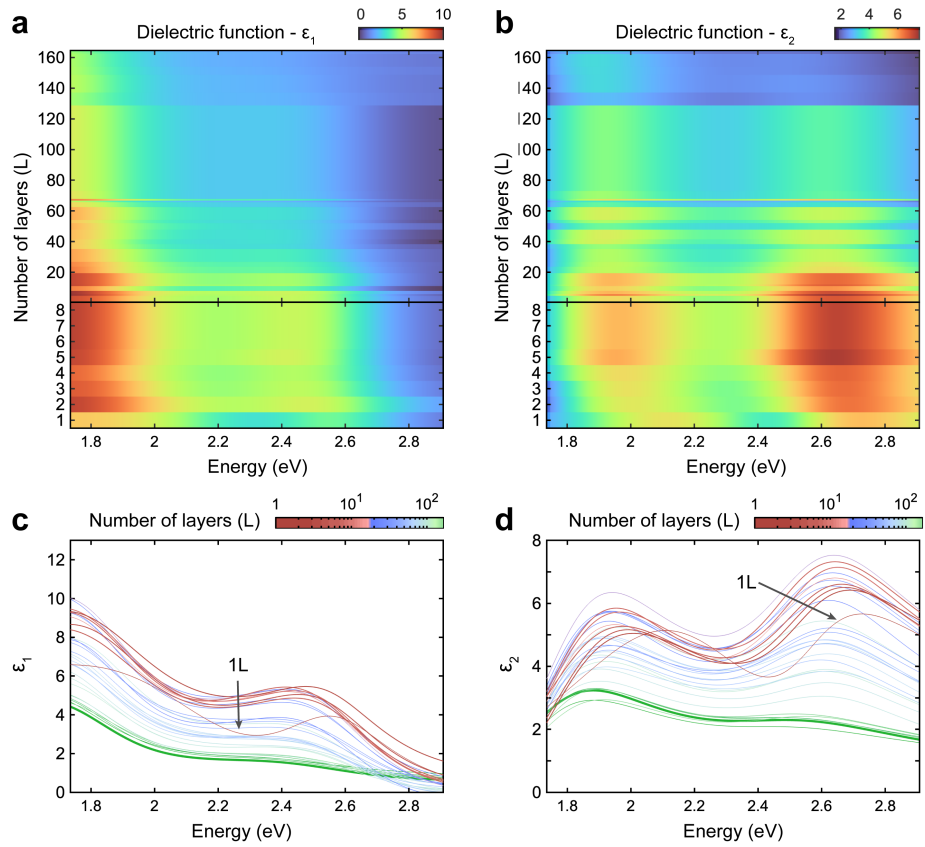

**Figure S9 | Map of the complex optical dielectric function of CrI$_3$ composed by independently fitting the transmittance spectra of samples with different thicknesses.** The calculated real ($\varepsilon_1$) and imaginary ($\varepsilon_2$) parts of $\tilde{\varepsilon}(\omega)$ are shown in panels (a) and (b), respectively. The computed $\varepsilon_1$ ($\hbar\omega$) (c) and $\varepsilon_2$ ($\hbar\omega$) (d) have been also plotted as a function of the excitation energy $\hbar\omega$ for different CrI$_3$ crystals of different thicknesses. The red traces correspond to thin layers (from 1L to 13L), the blue traces to layers ranging from 14L to ~100L and the green traces to bulk (up to 164L). This figure displays independent fits of specific thickness spectra in comparison to Figure 2 that shows a simultaneous fit to the complete layer-dependent dataset. A good agreement between the two figures can be observed.

# 4. Spatially resolved optical properties

Hyperspectral analysis allows to spatially resolve the optical properties of a 2D material. In Figure S10a we show the optical transmission image of different $CrI_3$ flakes with their respective thickness measured by AFM, while Figure S10b displays the spatially resolved intensity of the high energy transmittance dip of $CrI_3$ (at 2.66 eV). As for the peak position map of Figure 3 of the manuscript, this has been obtained by calculating pixel by pixel the corresponding transmittance spectrum and fitting its high energy dip position (Figure 3c) and transmittance intensity (Figure S10b), then ascribing these values to each pixel. Background pixels were automatically set to the minimum value. The color scale can also be compared to the values extracted from the analyzed transmittance spectra as a function of the number of layers (Figure S10c) showing a good agreement.

The spatial resolution of the optical properties of a 2D material results to be a very intuitive and powerful method to gain a fast insight on the properties of many different sample thicknesses at a same time. By tracking the peak transmittance intensity, as shown in Figure S10b, it is possible to gain a large contrast on the thinnest layers. Once transmittance intensity has been associated to the respective layers thickness characterized by AFM, this becomes a very powerful tool to quickly and reliably gain information on all flakes thickness of a whole image with a high spatial resolution. This process could be used for the automatic analysis of flakes thicknesses with clear advantages in terms of time saving (data images acquisition takes about 10 minutes for a whole zone) and non-invasive way, very relevant especially for air sensitive 2D materials.

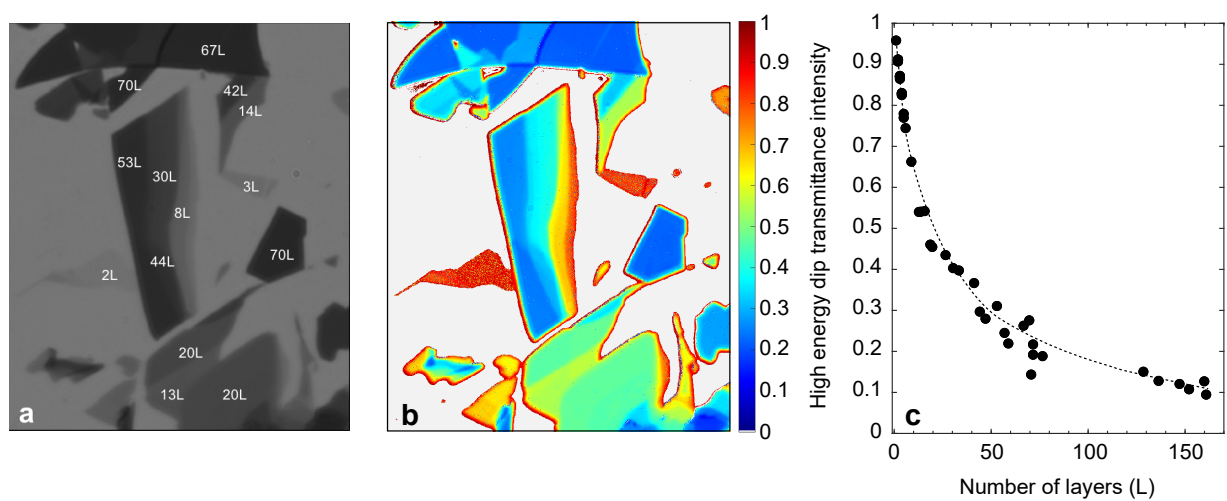

**Figure S10 | Spatially resolved layer dependence of the optical properties of $CrI_3$.** (a) Optical transmission image of $CrI_3$ flakes deposited on a glass slide. The number of layers of each section of the sample has been characterized by AFM and labeled on each flake in the image. (b) Spatially resolved map of zone shown in (a) displaying the transmittance value at the to display intensity at the high energy transmittance dip for each flake. This has been obtained by extracting the transmittance spectrum at each pixel, fitting the transmittance in the spectral vicinity of the high energy dip and obtaining the energy at the local minimum, and ascribing this value to each pixel. Background pixels where automatically set to 0. (c) Evolution with the number of layers of the minimum transmittance intensity at the high energy dip, obtained by relating analyzed transmission images to results obtained by AFM. Dots are experimental data while dotted line is guide to the eye. Once transmittance intensity dependence with layer thickness is calibrated, this technique allows to optically analyze thin layers thickness with a high throughput. This method is particularly convenient as it is fast and non-invasive.

# 5. Dependence of the complex dielectric function with layer number and interlayer distance

We carried out first-principles calculations on few-layer $CrI_3$, up to 10 L using the DFT+U+J scheme with U = 4.1 and J = 0.6 eV, as previously established for this material [7], based on the constrained random-phase approximation (cRPA) [8]. For all cases, we consider an ideal monoclinic stacking between adjacent layers and antiferromagnetic coupling.

Figure S11 shows the theoretically computed evolution of the real and imaginary parts of the dielectric function of $CrI_3$ with increasing number of layers for the full energy range calculated from 0 eV to 4 eV.

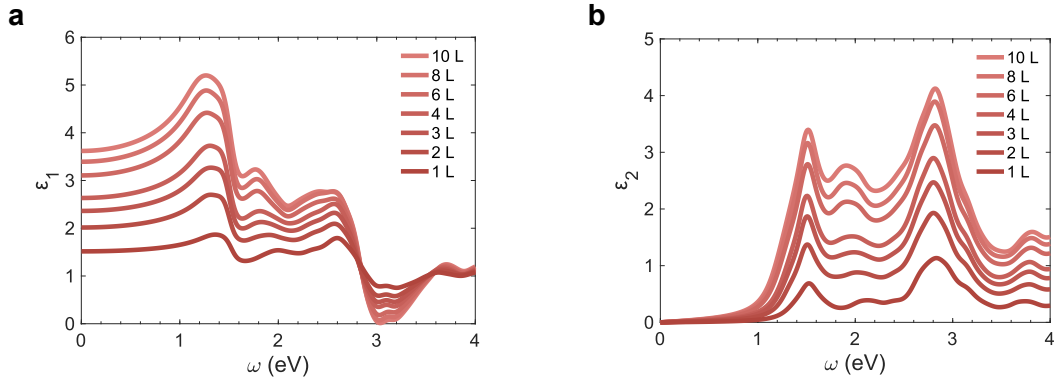

**Figure S11 | Evolution of the real and imaginary parts of the dielectric function of $CrI_3$ with increasing number of layers.** This calculation was performed in the z-direction perpendicular to the layers.

In Figure S12 we show the evolution of the intensity of the e2 peak at 1.5 eV for bilayer $CrI_3$ with the interlayer distance. The other peaks reported a similar trend. The increase in the inter-layer distance decreases the polarizability of the wave-functions in the direction perpendicular to the layers, also decreasing the magnitude of the complex dielectric function. These results clearly indicate that small changes in the crystal structure and stacking could eventually modify the intensity of the complex dielectric function. Considering that in layered bulk crystals the interlayer distance increases due to stronger van der Waals interactions and modifications in the crystal structure, our theoretical results give hints on the crossover and the different thickness regimes observed experimentally.

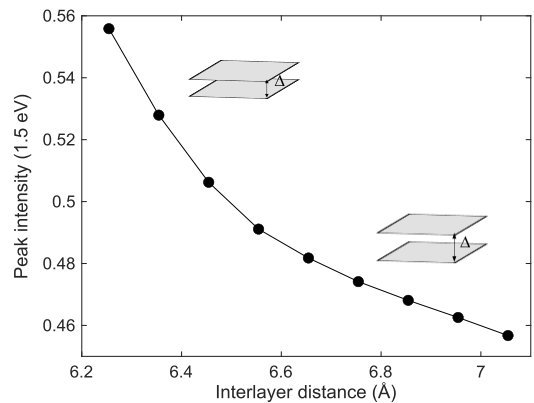

**Figure S12 | Dependence of the intensity of the 1.5 eV peak in the imaginary dielectric function with the inter-layer distance.**

| Number of layers | 2 | 4 | 6 | 8 | 10 | Bulk |
|---|---|---|---|---|---|---|
| Average Cr-Cr interlayer distance (Å) | 6.654 | 6.640 | 6.644 | 6.649 | 6.681 | 6.845 |

**Table S2 | Evolution of the Cr-Cr interlayer distance with the number of layers based on first principles calculations.**